\documentclass[amsmath,amssymb,superscriptaddress,twocolumn]{revtex4-2}
\usepackage[utf8]{inputenc}
\usepackage{hyperref}
\usepackage{graphicx}
\usepackage{dcolumn}
\usepackage{bm}
\usepackage{xcolor}
\usepackage[normalem]{ulem}

\begin{document}

\title{Devitrification and Melting Dynamics in Vapor Deposited Water Ice}

\author{Fabio Leoni}
\email{fabio.leoni@uniroma1.it}
\affiliation{Dipartimento di Fisica, Universit\`a degli Studi di Roma La Sapienza, Piazzale Aldo Moro 5, Rome, 00185, Italy}

\author{Fausto Martelli}
\affiliation{IBM Research Europe, Hartree Centre, Daresbury WA4 4AD, UK}
\affiliation{CNR-Istituto dei Sistemi Complessi, Piazzale Aldo Moro 5, Rome, 00185, Italy}
\affiliation{Dipartimento di Fisica, Universit\`a degli Studi di Roma La Sapienza, Piazzale Aldo Moro 5, Rome, 00185, Italy}

\author{John Russo}
\affiliation{Dipartimento di Fisica, Universit\`a degli Studi di Roma La Sapienza, Piazzale Aldo Moro 5, Rome, 00185, Italy}

\begin{abstract}
The equilibration dynamics of ultrastable glasses subjected to heating protocols has attracted recent experimental and theoretical interest. With simulations of the mW water model, we investigate the devitrification and melting dynamics of both conventional quenched (QG) and vapor deposited (DG) amorphous ices under controlled heating ramps. By developing an algorithm to reconstruct hydrogen-bond networks, we show that bond ring statistics correlates with the structural stability of the glasses and allows tracking crystalline and liquid clusters during devitrification and melting.
We find that QG melts in the bulk, whereas melting in DG preferentially begins near the free surface. During devitrification, the DG shows an excess of 5-membered rings near the free surface, which is consistent with its tendency to nucleate the crystal phase in this region.
Additionally, the DG shows an Avrami exponent exceeding the standard $1+$d behavior, while both glasses display the same sub-3d growth of liquid clusters across heating rates, indicating that the DG enhanced exponent stems from its higher kinetic stability.
\end{abstract}

\maketitle

\section{Introduction}

Since the experimental realization of ultrastable glasses (UGs) by vapor depositing molecules on a cold substrate \cite{swallen2007}, both experimental and computational studies have extensively explored the formation of UGs across various materials and model systems by tuning parameters such as the deposition rate and substrate temperature \cite{ediger2017,rodriguez2022}, and more recently the substrate softness \cite{luo2024}. 

One of the most striking features distinguishing UGs from conventional quenched glasses (QG)--which are produced by rapidly cooling the liquid melt to avoid crystallization--is their enhanced kinetic stability \cite{dalal2012,chua2015,rodriguez2015,whitaker2015,tylinski2016,fullerton2017,ediger2017,rodriguez2022}. This enhanced stability often manifests as a higher onset temperature $T_o$ upon heating, with $T_o$ the temperature at which the glass begins the transformation into the supercooled liquid, and has significant implications for technological applications where avoiding the deterioration of the glass properties over time is critical. In this work, we consider melting as the transition from a solid phase (either glassy or crystalline) to the liquid phase, while devitrification refers to the transformation from the glassy state into a crystalline phase, which can be mediated in some conditions by the formation of a supercooled liquid phase.

Only in recent years have experimental \cite{vila2023,ruiz2023} and computational \cite{flenner2019,herrero2023,chacko2024} studies begun to address the equilibration dynamics of UGs subjected to heating ramps. 
A pivotal experimental advancement by Ruiz-Ruiz et al.~\cite{ruiz2023} enabled real-time, microscopic-scale observation of melting in highly equilibrated glasses. 
Their results showed that melting in UGs proceeds heterogeneously in space, with sparse liquid-like regions separated by microscale distances (as also shown in Ref.~\cite{kearns2010} for indomethacin)—unlike QG, where such regions are spaced at the nanoscale and their transformation into supercooled liquid upon heating is found to proceed more homogeneously \cite{sepulveda2013}.
From a computational standpoint, studies using Lennard-Jones (LJ) polydisperse systems have revealed that the equilibration dynamics of melting UGs differ markedly from both equilibrium relaxation and aging dynamics \cite{herrero2023}. 
Deviations from classical Avrami kinetics \cite{fanfoni1998} of the melting clusters have been observed in Ref.~\cite{herrero2023}.
The increased stability of the glass, promoted by surface and subsurface mobility enhancement in vapor deposited materials \cite{swallen2007,stevenson2008,lyubimov2013,reid2016,berthier2017,samanta2019,ferron2022,leoni2023a}, has been shown to influence the crossover from bulk- to surface-mediated melting dynamics in both deposited and bulk UGs \cite{flenner2019}. 
Molecular dynamics (MD) simulations of a polydisperse LJ glass \cite{chacko2024} have shown that, even if the equilibration of the glass during heating proceeds by domain growth, there is no evidence to support a nucleation mechanism. 
Instead, melting has been found to proceed through local rearrangements which propagate mobility to neighborhoods \cite{chacko2024}, a process known as dynamic facilitation \cite{leonard2010,keys2011,sepulveda2013,herrero2024}.
The self-propagation of mobility can give rise to avalanches of plastic events \cite{candelier2010} or displacements correlated in space and time \cite{leoni2025}.
An Avrami-like domain growth during heating ramps can still be derived from a dynamical facilitation description \cite{chacko2024}.

In a recent study \cite{leoni2024}, the authors have simulated via MD the formation of DG of tetrahedral materials-- systems characterized by networks of strong directional bonds--using the generalized Stillinger-Weber (SW) potential \cite{stillinger1985}. 
This potential has been used to describe fundamental properties of several materials employed in the semiconductor and electronics industry such as germanium and silicon \cite{molinero2006}, and pure water \cite{smallenburg2014,russo2018}, especially in relation to ice nucleation and crystallization  \cite{moore2011,molinero2017,leoni2021}.
Results obtained in Ref.~\cite{leoni2024} demonstrated that DG in these systems can exhibit ultrastability, outperforming QG formed under equivalent thermodynamic conditions, within a temperature range constrained by crystal nucleation at higher temperatures.

In this study, we focus on the devitrification and melting dynamics of both deposited and quenched glasses during heating ramps applied at different rates. We use the SW potential with parameters tailored to describe water, reducing it to the coarse-grained monatomic water (mW) model \cite{molinero2009}. 
The study of amorphous water ices is of great interest in fields going from astrophysical and planetary science \cite{tonauer2023}, to cryopreservation of biological systems \cite{alba-simonesco2022}, to the understanding of water anomalies and phase diagram \cite{martelli2020}.

Given that many thermodynamic, dynamic, and structural properties of water are influenced by its hydrogen-bond network (HBN) \cite{martelli2019unravelling,russo2022}, we develop a novel algorithm to assign oxygen-hydrogen (OH) bonds between mW point molecules. This framework enables us to analyze the resulting HBN, including ring statistics of varying orders. 
Several properties of water at different conditions, from the liquid to the solid phases, have been linked to features of its topological bond network \cite{neophytou2022,gutierrez2024}.

Moreover, by characterizing this HBN, we can identify energetically neighboring particles and trace the formation and evolution of crystalline clusters during devitrification, as well as liquid clusters during melting. This approach provides a detailed microscopic view of the transformation dynamics and reveals the structural signatures underlying the enhanced stability of ultrastable glasses.


\section{Methods}

\subsection{System Model}
Molecules are modeled with the monatomic water model mW \cite{molinero2009}, which is described by the Stillinger-Weber \cite{stillinger1985} potential for a specific choice of the parameters.
This potential includes two-body ($\phi_2$) and three-body ($\phi_3$) terms in the following form:
\begin{equation}
E=\sum_i\sum_{j>i}\phi_2(r_{ij})+\sum_i\sum_{j\ne i}\sum_{k>j}\phi_3(r_{ij},r_{ik},\theta_{ijk})    
\end{equation}
where $r_{ij}$ is the distance between particle $i$ and $j$, and $\theta_{ijk}$ is the angle formed by the triplet of particles $ijk$. The strength of $\phi_3$, which favors the formation of the tetrahedral angle $\theta_0=109.47^o$, is controlled by the tetrahedral parameter $\lambda$.
$\sigma$ is the size of particles and the cutoff is $a=1.8\sigma$.
For the following choice of the parameters (appearing in $\phi_2$ and $\phi_3$), $\lambda=23.15$, $\epsilon=6.189$~kcal/mol and $\sigma=2.3925$~\AA, the SW model reduces to the mW \cite{molinero2009}.

\subsection{MD simulations}

Molecular dynamics (MD) simulations are performed with LAMMPS \cite{LAMMPS}. The system setup is the same as that described in Ref.~\cite{leoni2024}.
It consists of a simulation box with dimensions $L_x=L_y=15\sigma=35.89$~\AA, and $L_z$ big enough to accommodate all the particles deposited. Periodic boundary conditions are taken along $x$ and $y$.
The time step is set to $dt=5$~fs, as done in other deposition simulations of mW particles \cite{lupi2014,leoni2024}.
The properties of the systems are computed in the core defined by all the particles forming the QG or the DG excluding those in a $3\sigma$-thick layer in contact with the free surface and with the substrate.   

The glass is deposited by sequentially injecting individual particles from random positions at the top of the box onto the substrate maintained at a constant temperature $T$ (NVT ensemble), as detailed below.
Each particle is introduced with a fixed velocity component in the deposition direction, $v_z=0.01$~\AA/fs, and random velocity components in the transverse directions (i.e., x and y), where $v_x$ and $v_y$ range between -0.001 and 0.001~\AA/fs. This velocity distribution corresponds to a source temperature of approximately $\sim 1000$~K.
$N=5000$ particles are deposited with deposition rate $\gamma_{DG}=\Delta z/\Delta t$, where $\Delta z$ is the size of the deposited layer along $z$ and $\Delta t$ the elapsed time. 

The substrate composed of 500 particles is prepared by depositing them onto a system that already contains 500 randomly arranged particles of the same type.
This initial random configuration has a density $\rho=\rho(T_m)$, where $T_m$ is the melting temperature.
To avoid mixing between the deposited particles and those in the random configuration—particularly at temperatures above $T_g$—each particle in the random distribution is independently tethered to its original position using a spring force with stiffness $k_{rand}=1$.
The 500 particles intended to form the substrate are deposited onto this random layer at a rate of $\gamma_{sub}=5$~\AA/ns, using the same velocity vector as employed in the primary deposition simulation.
After deposition, this results in a substrate of 500 particles, to which a spring force of stiffness $k_{sub}=1$ is independently applied to each particle, following the same procedure used for the random distribution.

The conventional quenched glass (QG), which we compare to the vapor-deposited glass (DG) in the following section, is prepared using a protocol similar to the one described in Refs.~\cite{leoni2023a,leoni2024}.
The process involves taking a DG, melting it at a temperature $T_i>T_m$, and then equilibrating the system down to $T_f$ at a cooling rate $\gamma_{QG}=\Delta T/\Delta t = (T_i-T_f)/\Delta t$.
Here, $\Delta t$ denotes the time required to quench the system from $T_i$ to $T_f$, and for the chosen rate $\gamma_{QG}=10$~K/ns, this corresponds to integrating $5\cdot 10^6$ time steps.
The selected quench rate, $\gamma_{QG}=10$~K/ns, is sufficiently rapid to prevent crystallization in the mW model.
It is important to note that while in the case of DG the temperature $T$ refers to that of the substrate (which is in contact with the thermal bath), for the QG, the temperature corresponds to that of the entire system.

\subsection{Hydrogen bond network (HBN) algorithm}

To build the hydrogen bond network associated with the mW monatomic water model, we use the algorithm described by the following steps:
\begin{enumerate}
    \item find the 4 nearest neighbors of each particle $i$ as the neighbors with the lowest bond energy;
    \item compute the tetrahedral order parameter $q_i$ for each particle as:
\begin{equation}
q_i=1-\dfrac{3}{8}\sum_{j=1}^{3}\sum_{k=j+1}^{4}\left(cos\theta_{ijk}+\dfrac{1}{3}\right)^2    
\end{equation}
where $j$ and $k$ are indices going from 1 to 4, indicating the 4 nearest neighbors of particle $i$. $\theta_{ijk}$ is the angle formed between vectors $r_{ij}$ and $r_{ik}$. $\langle q_i\rangle=0$ for a random distribution of the 4 nearest neighbors of $i$, and $\langle q_i\rangle=1$ for a perfectly tetrahedral configuration.
    \item Look for non-reciprocal bonds (they should be associated with either acceptors or donors): if $q_i>q_j$ then add a bond between oxygen $j$ and oxygen $i$, while if $q_j>q_i$ then remove the bond between oxygen $i$ and oxygen $j$.  
\end{enumerate}

\subsection{Melting}

To identify liquid particles at time $t$ we compute how many near neighbors of each particle have changed from the initial time $t_0$, indicated with $C(t,t_0)$\cite{herrero2023,sarkar2025}. When $C$ reaches the value 0.5 we consider a particle to be in the liquid phase. We verified that the results obtained in this work are robust against a variation in the threshold of $C$.

To characterize structural properties of liquid clusters during melting, we compute the radius of gyration of each cluster as $R_g=\sqrt{Tr(S)}$, where $S$ is the gyration tensor and $Tr$ is the trace of the associated matrix which can be written as
\begin{equation}
S_{\alpha\beta}=\dfrac{1}{2n_c^2}\sum_{i=1}^{n_c}\sum_{j=1}^{n_c}(r_{\alpha}^{i}-r_{\alpha}^{j})(r_{\beta}^{i}-r_{\beta}^{j})     
\end{equation}
where $\alpha,\beta=x,y,z$, and $r_{\alpha}^i$ is the $\alpha$ component of the position vector of particle $i$.


\section{Results}

\subsection{Ring statistics and thermodynamic stability}

We begin by analyzing the structural characteristics of the HBN in both QG and DG by computing the fraction of bond rings of various degrees. Here, the degree refers to the number of edges (or bonds) forming a closed loop. 
\begin{figure}[ht!]
\begin{center}
\includegraphics[clip=true,width=8.5cm]{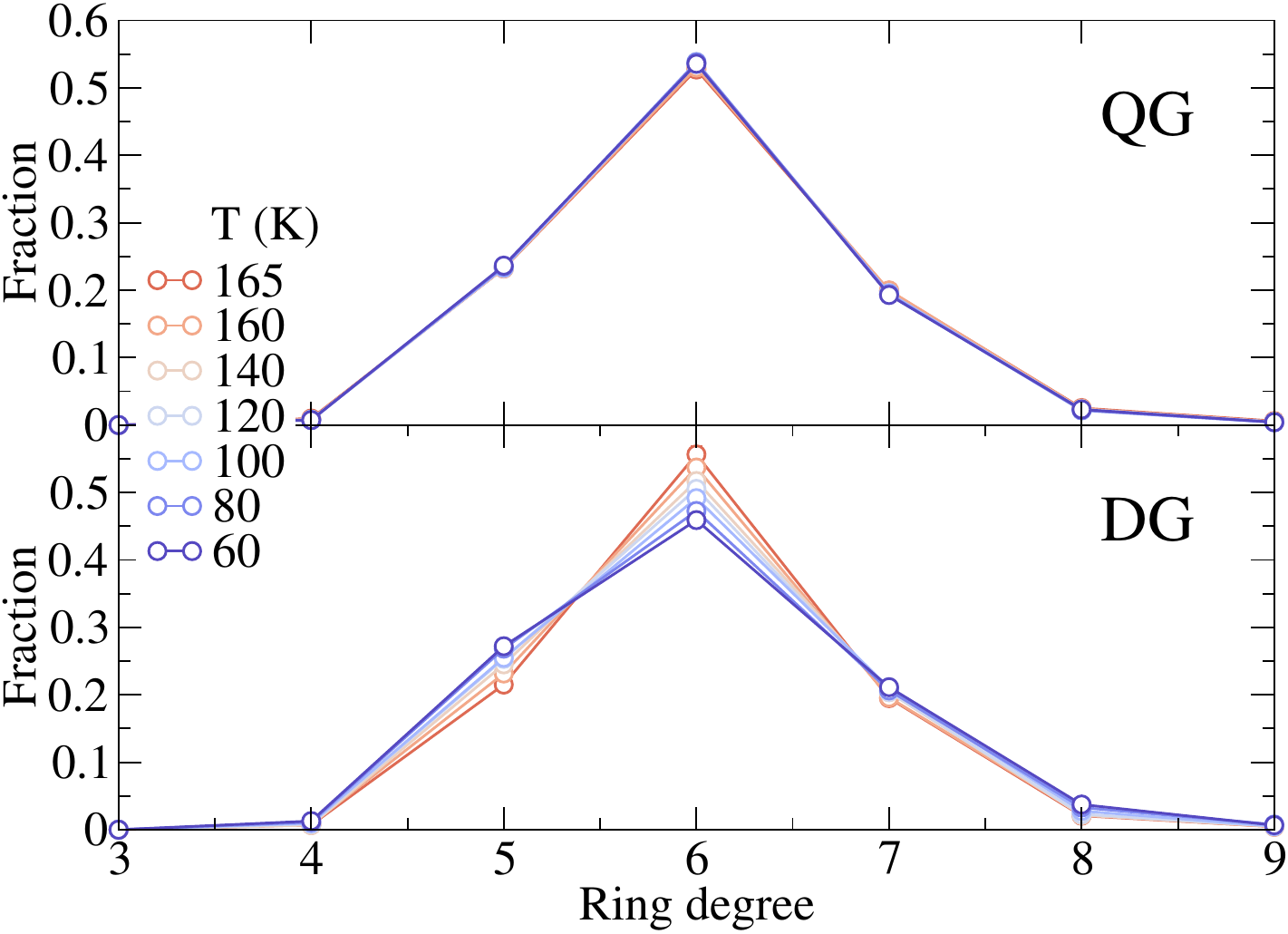}
\end{center}
\caption{Fraction of rings versus rings degree present in the core of the QG (upper panel) and the DG (lower panel) at different quench and substrate temperatures, respectively. The quench rate is $\gamma_{QG}=10$~K/ns with a starting and final temperature of $T_i=300$~K and $T_f=T$ (as indicated in the legend), respectively. For the DG, the deposition rate is $\gamma_{DG}=2.5$~\AA/ns.}
\label{fig:Fig1_2}
\end{figure}
We first consider the QG obtained via rapid quenching at a rate of $\gamma_{QG}=10$~K/ns and the DG formed via vapor deposition at rate $\gamma_{DG}=2.5$~\AA/ns, across a range of temperatures (see Figs.~\ref{fig:Fig1_2} and Fig.~\ref{fig:Fig3}(a)). These results are compared with the thermodynamic stability, as captured by the potential energy per particle (Fig.~\ref{fig:Fig3}(b)). 
\begin{figure}[t!]
\begin{center}
\includegraphics[clip=true,width=8.5cm]{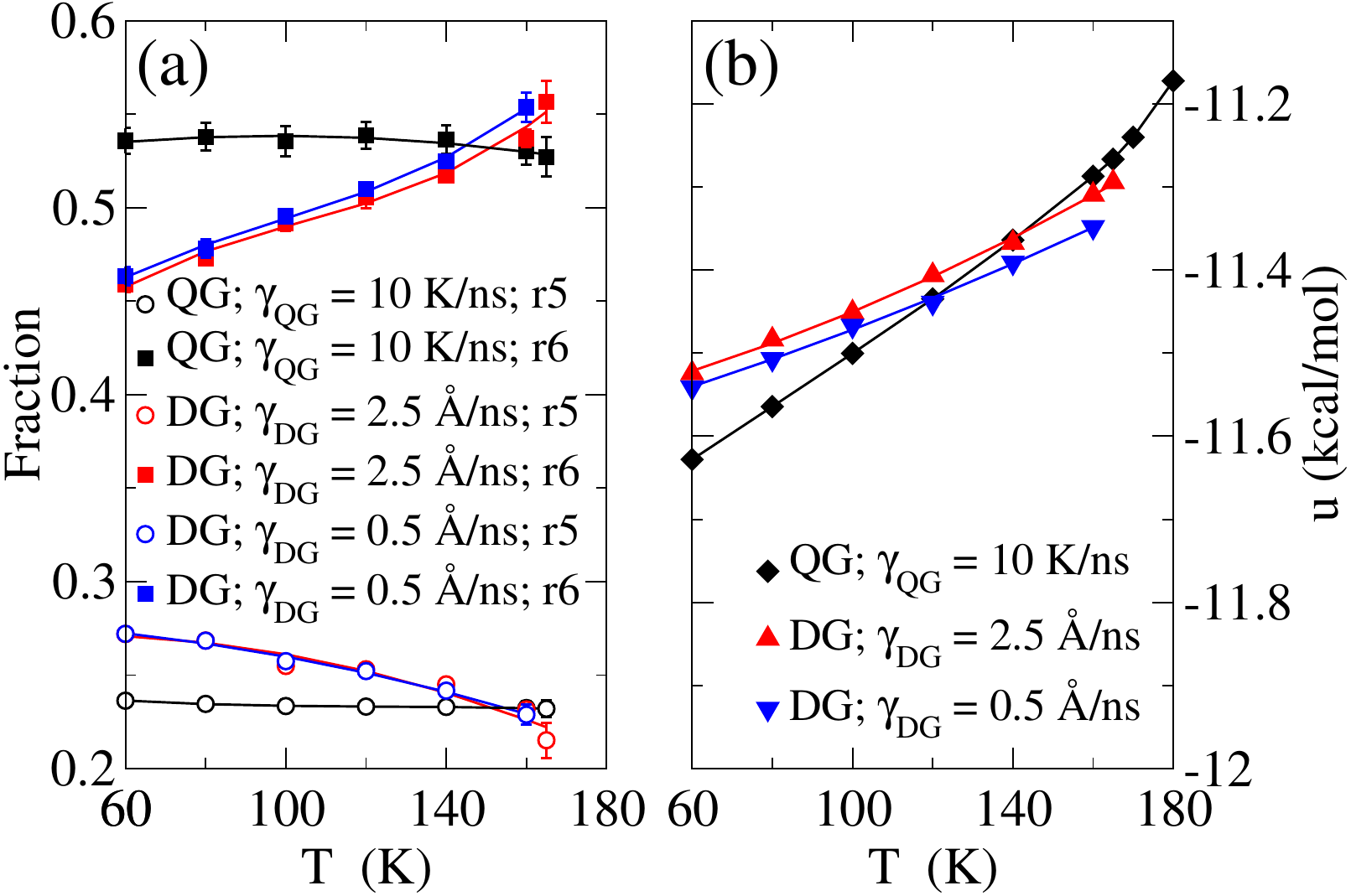}
\end{center}
\caption{Panel (a): fraction of rings of degree 5 (r5) and 6 (r6) versus temperature for the QG obtained upon quenching at $\gamma_Q=10$~K/ns, and the DG deposited at $\gamma_D=0.5$ and 2.5~\AA/ns. Panel (b): Potential energy per particle for QG and the two DGs considered.}
\label{fig:Fig3}
\end{figure}

The temperature range examined lies below the glass transition temperature $T_g\simeq 185$~K, where the deposited glass remains in a glassy state. This range also includes the window where the DG exhibits greater stability than the QG before crystallization becomes dominant at higher temperatures. 
Fig.~\ref{fig:Fig1_2}, upper panel, shows that the fraction of rings of different degrees in the QG are frozen at all temperatures, as expected for $T<T_g$. In contrast, from the lower panel of Fig.~\ref{fig:Fig1_2} we can see that the DG exhibits an increased fraction of rings of degree 6 and a corresponding decrease in rings of degree 5 as temperature increases, with other ring populations remaining relatively constant.
In Fig.~\ref{fig:Fig3}(a) we compare the ring statistics of the QG for $\gamma_{QG}=10$~K/ns with the DG for $\gamma_{DG}=0.5$ and $2.5$~\AA/ns. From it we can see that the fraction of 6-member rings correlates with the thermodynamic stability of the glasses as captured by the potential energy per particle versus temperature (Fig.~\ref{fig:Fig3}(b)).

\subsection{Devitrification under controlled heating ramps}

After considering structural properties of the different glasses obtained at a specific temperature, we perform heating ramps of the QG obtained at quench rate $\gamma_{QG}=10$~K/ns and the DG obtained at deposition rate $\gamma_{DG}=0.5$~\AA/ns using different heating rates $\gamma_H$. We are interested, on one side, in studying the devitrification process, and on the other side in the melting phenomenology in both glasses.
To investigate devitrification dynamics, we perform heating ramps on QG and DG samples at two heating rates, $\gamma_H=1$~and $10$~K/ns, starting from $T_i=160$~K (where DG is more stable than QG, but still glassy) up to $T_f=300$~K (well above the melting point $T_m\simeq 273$~K). 
We consider two heating rates to observe the different devitrification pathways expected for low and high $\gamma_H$. 
Indeed, at low $\gamma_H$, molecules have enough time to rearrange into a crystalline structure. On the other hand, at high $\gamma_H$, the glass is brought above $T_g$ quickly, leading to a supercooled liquid before crystallization begins.
These two different behaviors are observed for several glassy materials \cite{louzguine2007,herrero2023}, including water \cite{bhat2005,kringle2020}.

\begin{figure}[t!]
\begin{center}
\includegraphics[clip=true,width=8.5cm]{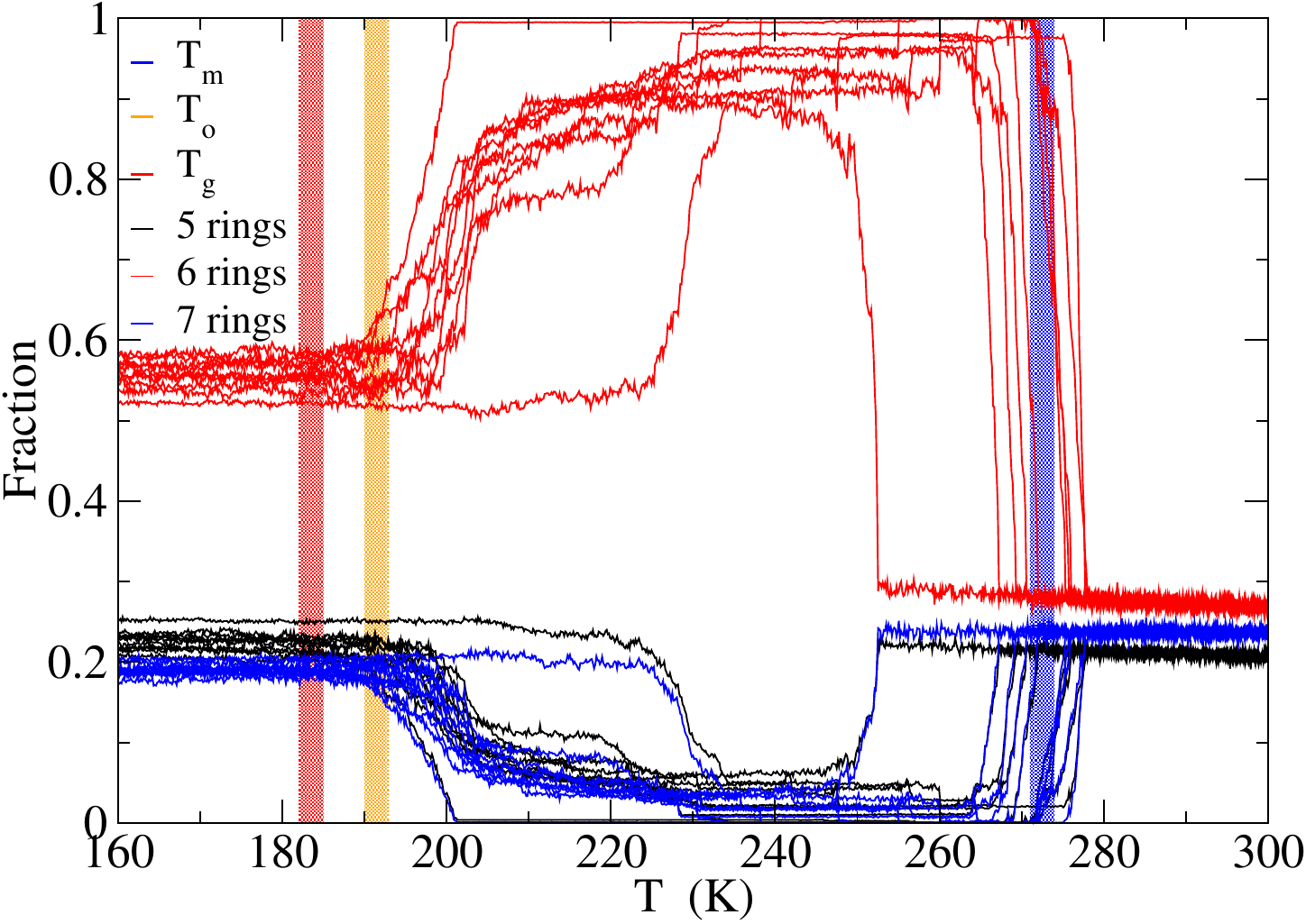}
\end{center}
\caption{Fraction of rings of degree 5, 6, and 7 versus temperature for 10 different samples of DG deposited at $\gamma_{DG}=0.5$~\AA/ns and heated from $T_i=160$~K to $T_f=300$~K with rate $\gamma_H=1$~K/ns. The vertical stripes indicate the onset temperatures.}
\label{fig:Fig5}
\end{figure}

\begin{figure}[ht!]
\begin{center}
\includegraphics[clip=true,width=8.5cm]{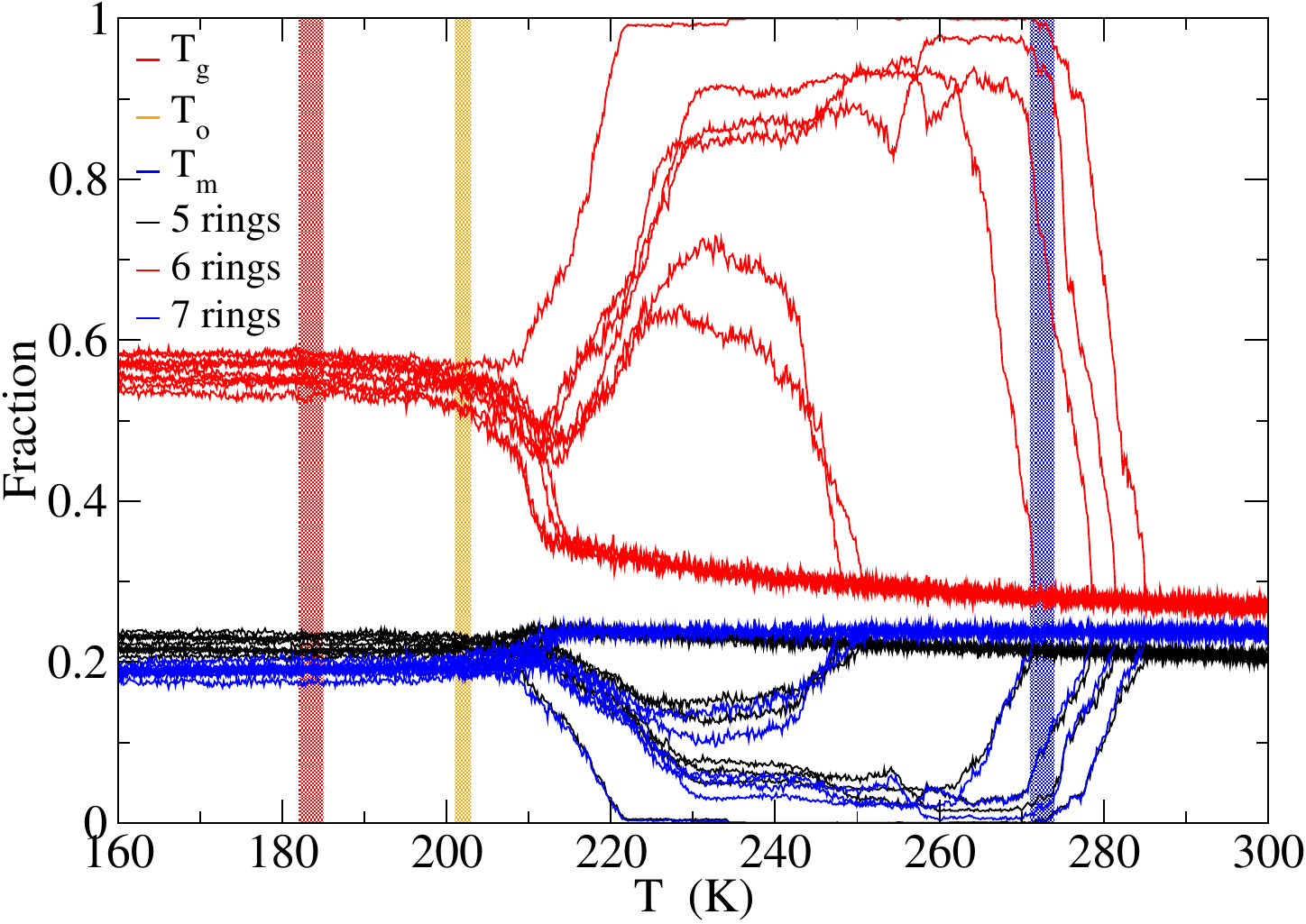}
\end{center}
\caption{Fraction of rings of degree 5, 6, and 7 versus temperature for 10 different samples of DG deposited at $\gamma_{DG}=0.5$~\AA/ns and heated from $T_i=160$~K to $T_f=300$~K with rate $\gamma_H=10$~K/ns. The vertical stripes indicate the onset temperatures.}
\label{fig:Fig7}
\end{figure}

For both glasses under all conditions and heating rates, we analyze 10 independent samples.
We compute the evolution with temperature of the fraction of rings of order 5, 6, 7 for all samples heated at rate $\gamma_H=1$~K/ns for the QG (not shown) and the DG (Fig.~\ref{fig:Fig5}), and rate $\gamma_H=10$~K/ns for the QG (not shown) and the DG (Fig.~\ref{fig:Fig7}).  

\begin{figure}[t!]
\begin{center}
\includegraphics[clip=true,width=8.5cm]{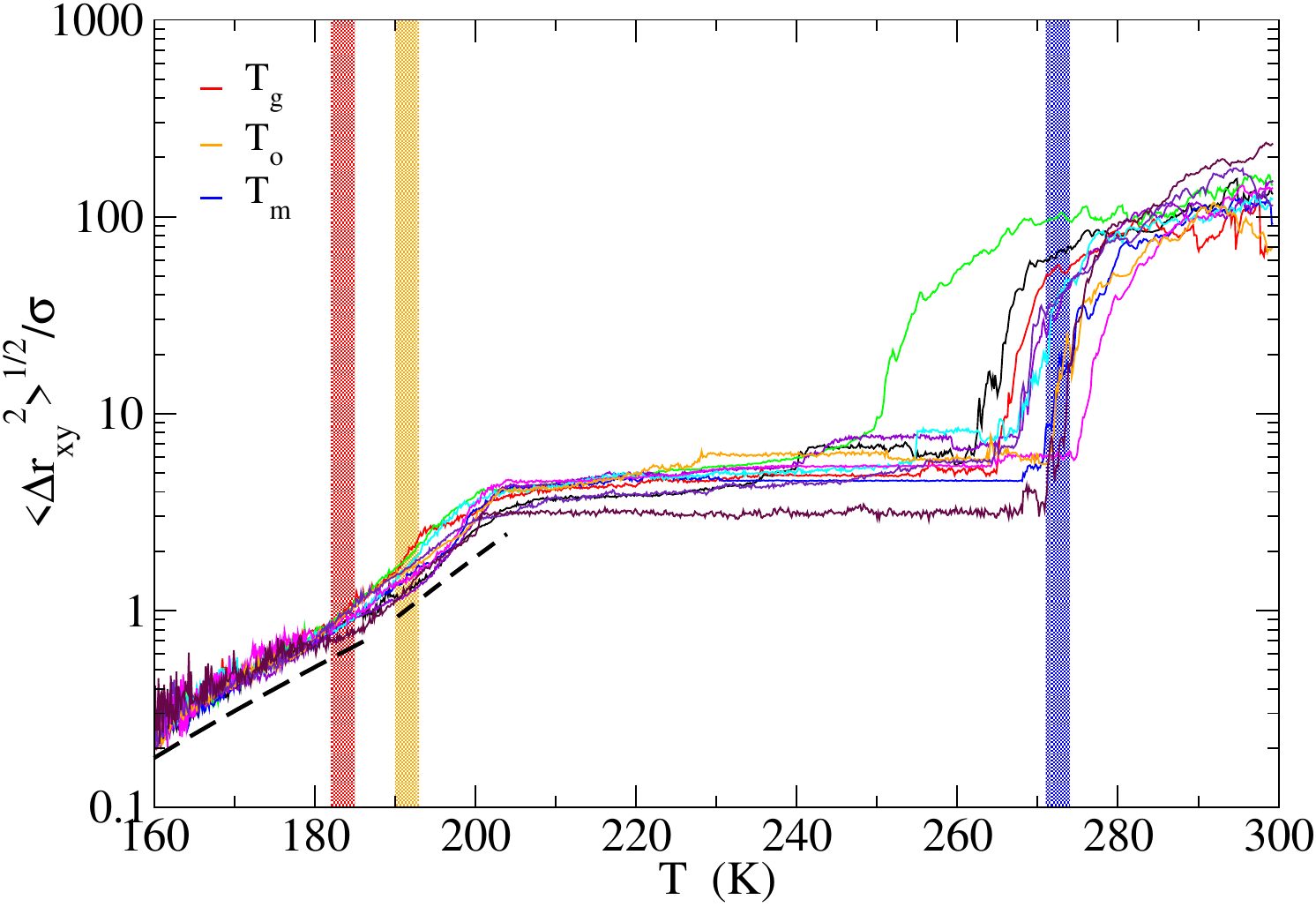}
\end{center}
\caption{Mean square displacement of 10 different samples of DG deposited at $\gamma_{DG}=0.5$~\AA/ns and heated from $T_i=160$~K to $T_f=300$~K with rate $\gamma_H=1$~K/ns. The black dashed lines show the slope of the MSD vs T in the glassy phase. Dashed lines are a guide for the eyes. The vertical stripes indicate the onset temperatures.}
\label{fig:Fig8}
\end{figure}

\begin{figure}[t!]
\begin{center}
\includegraphics[clip=true,width=8.5cm]{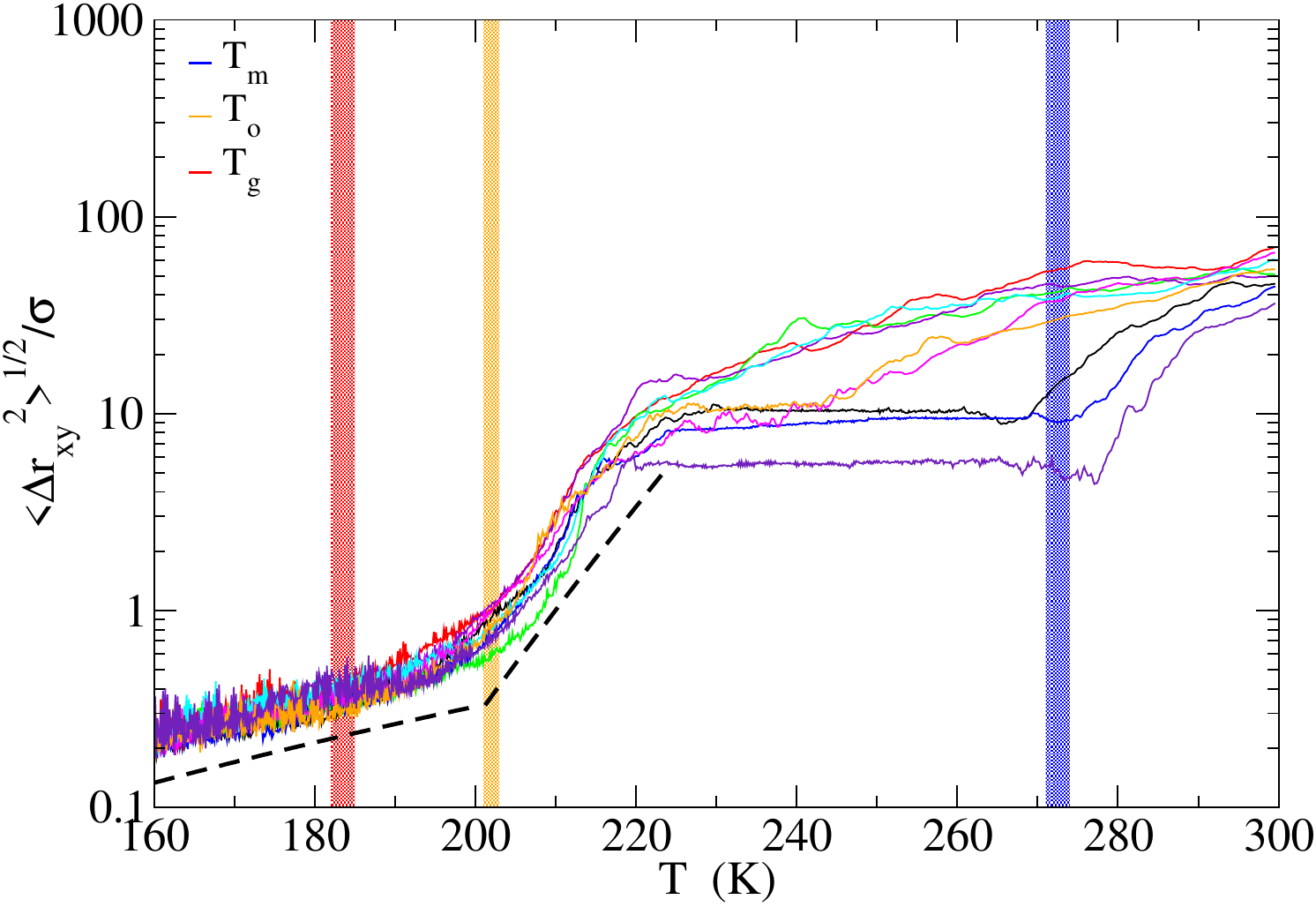}
\end{center}
\caption{Mean square displacement of 10 different samples of DG deposited at $\gamma_{DG}=0.5$~\AA/ns and heated from $T_i=160$~K to $T_f=300$~K with rate $\gamma_H=10$~K/ns. The black dashed lines show the slope of the MSD vs T. The black dashed lines show the slope of the MSD vs T in the glassy and supercooled phases. Dashed lines are a guide for the eyes. The vertical stripes indicate the onset temperatures.}
\label{fig:Fig9}
\end{figure}

For $\gamma_H=1$~K/ns, both QG and DG devitrificate from the glassy phase, starting to form crystallites at about the onset temperature $T_o$ \cite{leoni2024}, with $T_o(\gamma_H=1$K/ns$)\simeq 192$~K (indicated with a bold vertical orange bar in the figures). In contrast, at rate $\gamma_H=10$~K/ns, devitrification occurs only above $T_o(\gamma_H=10$K/ns$)\simeq 202$~K (indicated with a bold vertical orange bar in the figures).
A relevant difference between the devitrification process undergone at the two different heating rates lies in the emergence of an intermediate supercooled liquid phase between the glassy and the crystalline phase observed only for $\gamma_H=10$~K/ns.
This intermediate phase is characterized by a drop in the fraction of 6-member rings when the temperature exceeds $T_o(\gamma_H=10$K/ns$)$, followed by a sharp increase, indicating crystallization, except for 3 out of 10 samples which melt without forming crystals. 
The mean square displacement (MSD) further supports this picture. 
Indeed, for the glass heated at the low rate $\gamma_H=1$~K/ns, the MSD does not show a significant change in slope before crystallization (starting when the MSD flattens, indicating arrested dynamics, see Fig.~\ref{fig:Fig8} for the DG). 
On the other hand, for the glass heated at rate $\gamma_H=10$~K/ns, the presence of the supercooled phase corresponding to the drop in the fraction of 6-member rings above $T_o$ is confirmed by the change in slope appearing at about $T_o$, and signaling enhanced mobility in the supercooled liquid (see Fig.~\ref{fig:Fig9} for the DG).

\begin{figure}[ht!]
\begin{center}
\includegraphics[clip=true,width=8.5cm]{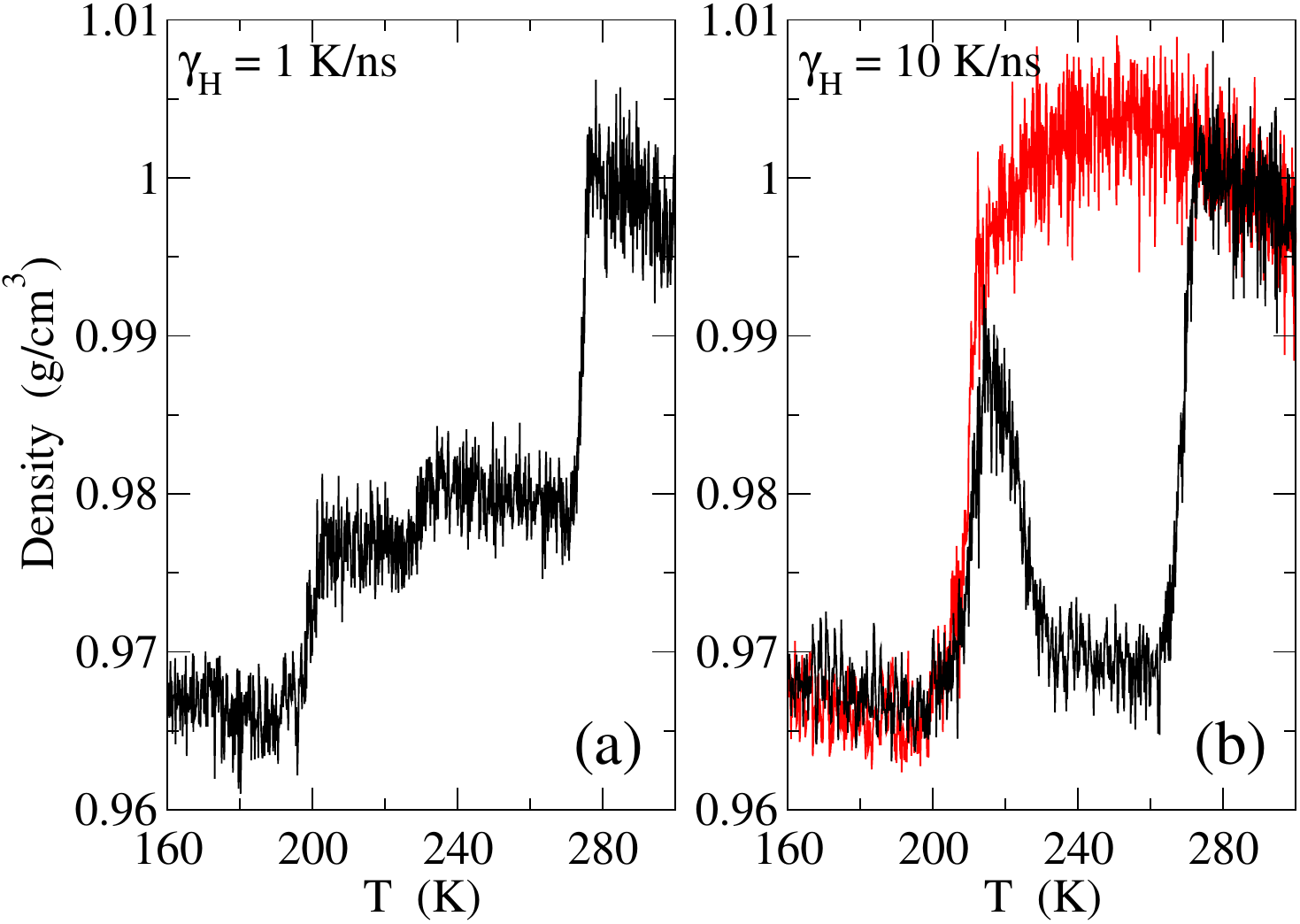}
\end{center}
\caption{Density versus temperature of typical samples of DG deposited at $\gamma_{DG}=0.5$~\AA/ns and heated at rate $\gamma_H=1$~K/ns (a) and $\gamma_H=10$~K/ns (b). At low heating rate (a), all samples crystallize and only one of them is shown. At high heating rate (b), we show a typical crystallizing (in black) and melting without forming crystallites (in red) samples.}
\label{fig:Fig11}
\end{figure}

Density profiles versus temperature also highlight this difference: the DG heated at $\gamma_H=10$~K/ns (see black curve in Fig.~\ref{fig:Fig11}(b)) that eventually crystallize show a density peak near $T=215$~K before dropping as crystallization proceeds. This feature is absent at the slower heating rate $\gamma_H=1$~K/ns (see Fig.~\ref{fig:Fig11}(a)).

Note that in water the liquid phase is denser than the solid phase. In Fig.~\ref{fig:Fig11}, liquid water at zero pressure reaches the typical density $\rho\simeq 1.0$~g/cm${^3}$, while the glassy phase at low temperatures assumes the typical low-density amorphous (LDA) phase density $\rho\simeq 0.96-0.97$~g/cm${^3}$.

\subsection{Melting dynamics and kinetic stability}

\begin{figure}[h!]
\begin{center}
\includegraphics[clip=true,width=8.5cm]{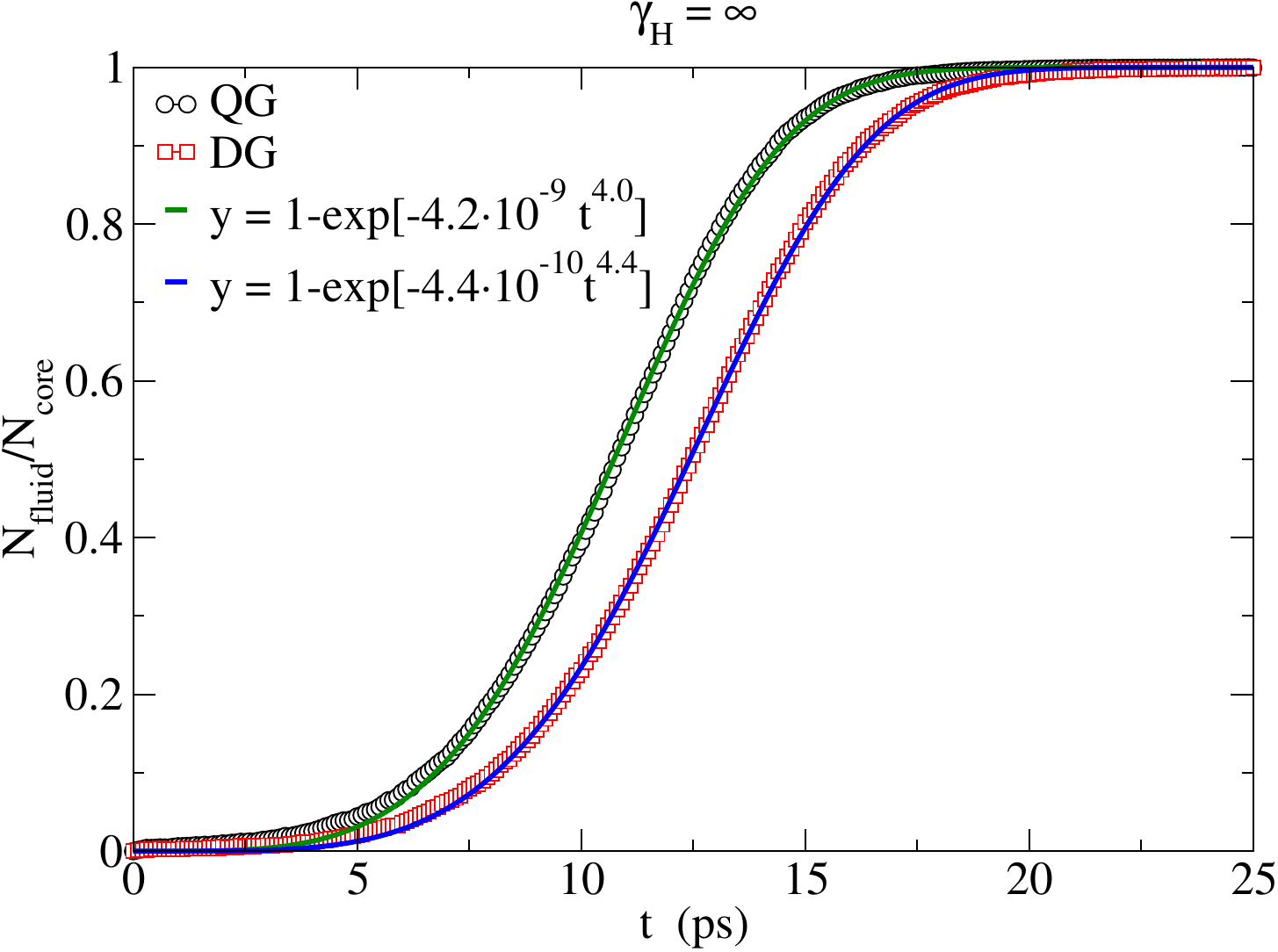}
\end{center}
\caption{Average fraction of fluid particles versus temperature for the QG and DG at infinite heating rate. The Avrami equation accurately fits the QG and DG data with an Avrami exponent of $4.0$ and $4.4$, respectively.}
\label{fig:Fig13}
\end{figure}

To probe melting behavior, we heat both QG and DG samples from $T_i=160$~K to $T_f=300$~K at the finite rates $\gamma_H=1$,~10~K/ns, and instantaneously (i.e., at infinite rate, $\gamma_H\rightarrow\infty$). A particle is identified as liquid if at least half of its neighbors differ from those at the beginning of the heating ramp (see Methods). 
The fraction of fluid particles $N_{fluid}/N_{core}$ (with $N_{core}$ the number of particles in the core, defined in Methods, and $N_{fluid}$ the particles of the core considered  fluid) during heating ramps at finite (
not shown) and infinite (Fig.~\ref{fig:Fig13}) rate shows the higher kinetic stability of the DG with respect to the QG. 

At infinite heating rate, since the heating is instantaneous and then the melting rate is constant to a good approximation, the melting curves can be fitted with the Avrami law~\cite{herrero2023}: $N_{fluid}/N_{core}=1-\exp(-Kt^{\alpha})$, with the time exponent predicted to be $\alpha=d+1=4$ in three dimensions (3d). 
From this fit, we obtain the exponents $\alpha_{QG}\simeq4.0$ and $\alpha_{DG}\simeq 4.4$.
From arguments involving the different growing mechanisms of liquid clusters in conventional and ultrastable glasses under instantaneous heating ramps, it is expected \cite{herrero2023} that, while the conventional glass, like the QG, should follow the Avrami law with exponent $1+d=4$ in 3d, the ultrastable one should follow it with a larger value of the exponent. The value of $\alpha_{DG}\simeq4.4$ we find in the current system is indicative of a distinct melting mechanism and a hallmark of ultrastability in the DG.

\subsection{Glass melting and liquid clusters growth}

\begin{figure}[t!]
    \centering
    \includegraphics[width=0.32\linewidth]{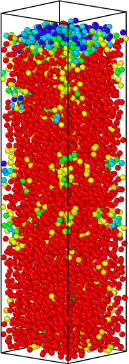}
    \includegraphics[width=0.32\linewidth]{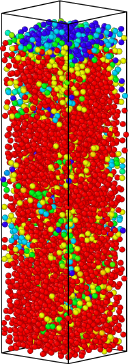}
    \includegraphics[width=0.32\linewidth]{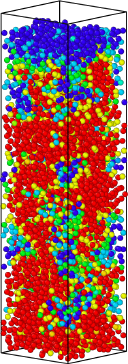}
    \caption{Melting of the DG deposited at rate $\gamma_{DG}=0.5$~\AA/ns and heated from $T_i=160$~K to $T_f=300$~K at rate $\gamma_H=10$~K/ns. From left to right, snapshots are taken at temperature $T=180, 190, 200$~K. Colors correspond to the value of $C$ (see Methods) going from 0 (red) to 1 (blue). Snapshots are made with Ovito \cite{stukowski2009}.}
    \label{fig:Fig15}
\end{figure}

\begin{figure}[h!]
    \centering
    \includegraphics[width=0.32\linewidth]{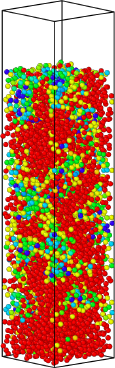}
    \includegraphics[width=0.32\linewidth]{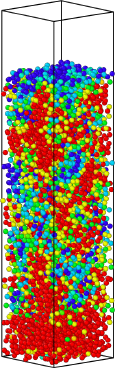}
    \includegraphics[width=0.32\linewidth]{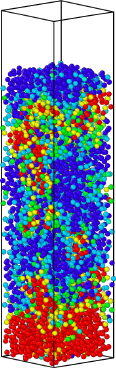}
    \caption{Melting of the QG quenched at rate $\gamma_{QG}=10$~K/ns and heated from $T_i=160$~K to $T_f=300$~K at rate $\gamma_H=10$~K/ns. From left to right, snapshots are taken at temperature $T=180, 190, 200$~K. Colors correspond to the value of $C$ (see Methods) going from 0 (red) to 1 (blue).}
    \label{fig:Fig16}
\end{figure}

Differences in the melting dynamics between QG and DG can be observed from snapshots of configurations taken at different temperatures for $\gamma_H=10$~K/ns with particles colored according to the value of $0\leq C\leq 1$ (see Methods). 
From snapshots (see Figs.~\ref{fig:Fig15}, \ref{fig:Fig16}) we can see that, while for the QG melting initiates uniformly throughout the sample, for the DG melting preferentially begins at the free surface, consistent with the heterogeneous transformation of stable glasses into liquids beginning at the surface of the sample observed for different materials heated at high rate \cite{swallen2009,ediger2017,kaur2023}.

\begin{figure}[ht!]
\begin{center}
\includegraphics[clip=true,width=8.5cm]{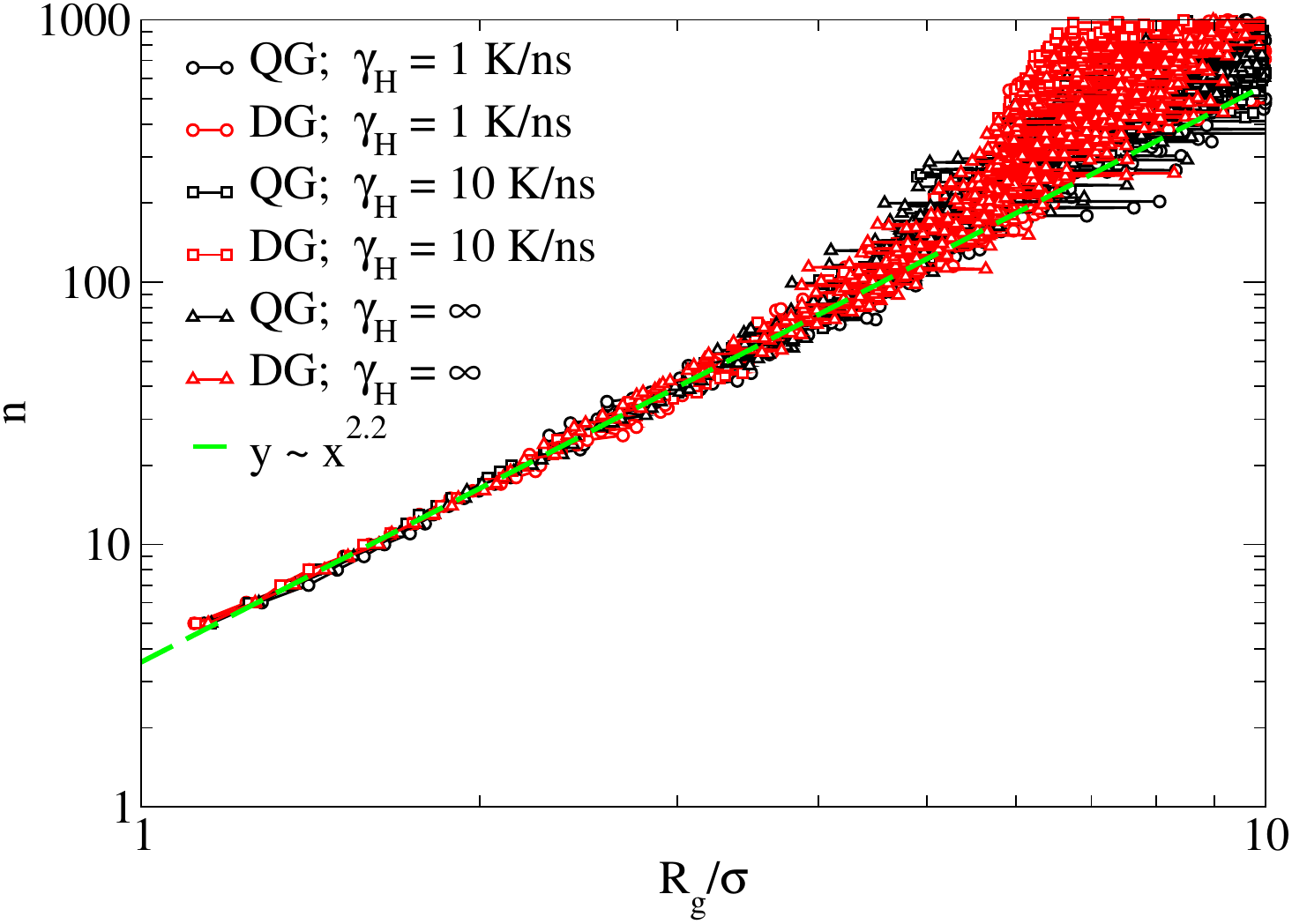}
\end{center}
\caption{Average size of fluid clusters $n$ versus their scaled Radius of gyration $R_g/\sigma$ for the QG and DG at different heating rates. The growth of the fluid clusters follows a sub-3D evolution with $n\sim R_g^{2.2}$.}
\label{fig:Fig17}
\end{figure}

We further characterize the spatial extent of liquid regions by computing the radius of gyration $R_g$ of liquid clusters (see Methods). 
In Fig.~\ref{fig:Fig17}, we show the size $n$ versus the normalized radius of gyration $R_g/\sigma$ for the QG and DG at all heating rates considered in this work, i.e., for $\gamma_H=1, 10, \infty$~K/ns.
In all cases, clusters follow the sub-3d growth law: $n\sim R_g^{2.2}$, indicating a common geometric growth regime. 
The enhanced Avrami exponent observed for DG, therefore, originates from its higher kinetic stability rather than a fundamentally different cluster growth geometry.


\subsection{Crystallization}

Due to its tendency to spontaneously form crystalline structures within accessible simulation timescales, the mW water model has been effectively used to describe fundamental aspects of ice nucleation and growth \cite{moore2011,molinero2017,leoni2021}, especially at high supercooling conditions \cite{espinosa2016,sanchez2022}. During heating ramps conducted at finite rates, most of the ten simulated samples of both the QG and DG exhibit crystallization via similar pathways and dynamics, with a few noteworthy differences. Specifically:

{\it i)} Multiple grains of randomly stacked cubic (Ic) and hexagonal (Ih) ice compete during the crystallization process (we emphasize here that geometric arguments favor polymorphism selection in molecular models of water towards Ih~\cite{martelli2022signatures}, as observed experimentally);
{\it ii)} The non-equilibrium melting temperature of crystalline clusters—ranging from 245 K to 277 K—depends strongly on cluster size and the presence of defects;
{\it iii)} Crystal melting generally initiates in regions with a high defect concentration, though in the absence of such regions, it typically proceeds symmetrically from the upper and lower surfaces (along the z-direction) of the cluster.
Differences between the QG and DG primarily relate to the spatial location of the nucleating clusters. While nucleation in QG samples can occur throughout the bulk of the layer, approximately one-third of DG samples show a preference for nucleation near the free surface.

The role of a free surface in influencing the ice nucleation pathway remains an open question \cite{tabazadeh2002,duft2004,gurganus2011}. It is still under debate whether nucleation preferentially occurs near the free surface \cite{sun2024} or within the bulk \cite{haji2017}, and whether such behavior is dependent on the model used in MD simulations \cite{li2013,haji2014,haji2017,sun2024}. A recent work \cite{sun2024} has highlighted the critical importance of 5-membered rings (associated with precursor structures, such as ice 0-like environments \cite{russo2014,leoni2019,leoni2021}) in the crystallization process.
Ref.~\cite{sun2024} reports an excess of 5-membered rings near the free surface in MD simulations using the TIP4P/Ice water model \cite{abascal2005}, which is shown to produce nucleation pathways similar to those observed in the mW model.
\begin{figure}[t!]
\begin{center}
\includegraphics[clip=true,width=8.5cm]{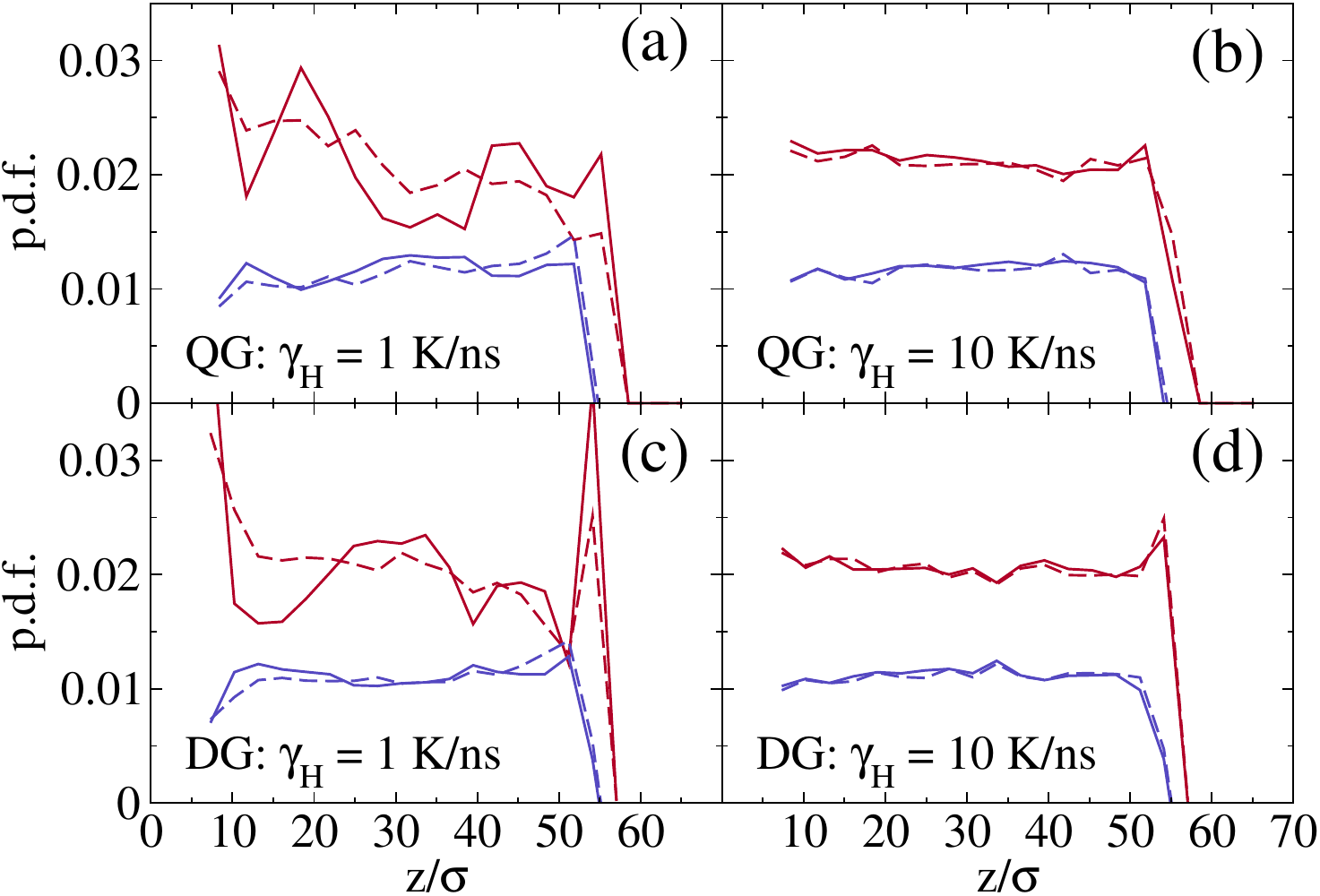}
\end{center}
\caption{Probability distribution function along the z direction of 5- (red) and 6-membered (blue) rings for the QG heated at rate (a) $\gamma_H=1$ and (b) 10~K/ns. Similarly for the DG heated at rate (c) $\gamma_H=1$ and (d) 10~K/ns. For a better readability, 6-membered rings curves are shifted along y of -0.01. Dashed and continuous lines refer to averages over a range of temperature of 196-202 K and 202-208 K, respectively.}
\label{fig:Fig18}
\end{figure}

Thanks to the algorithm we adopted to build the HBN from the mW point molecules configuration (see Methods), we can also study the evolution of the probability distribution function (p.d.f.) of rings of various orders across the deposition direction in both QG and DG samples for temperatures before crystallization takes place.
As shown in Fig.~\ref{fig:Fig18}, while it is difficult to establish the presence of an excess of ice-0-like structures (rich in 5-membered rings) near the free surface of the QG (a, b), our data clearly indicate this enhancement of 5-membered rings near the free surface in DG samples (c, d).

Given the limited size of our dataset (10 independent configurations), we cannot give a conclusive answer to the general problem of the ice nucleation pathway followed in the presence of a free surface. However, in agreement with Ref.~\cite{sun2024}, we definitively
observe an increase in 5-membered ring precursors, along with a nucleation preference, in DG samples near the free surface. Although not the central focus of this study, these findings highlight the relevant role of stability in the pathway to nucleation in presence of a free surface worth of further investigation.

\section{Conclusions}

In this work, we employed molecular dynamics simulations to investigate the differences in equilibration dynamics between conventional (quenched) and ultrastable (deposited) glasses subjected to thermal protocols. We considered a coarse-grained monatomic water model including three-body interactions, representative of a broad class of tetrahedral materials. 
We introduced a novel algorithm to build the hydrogen bond network from coarse-grained point-molecule configurations. This approach enabled us to analyze both the structural properties of the HBN, such as ring statistics, and the evolution of melting clusters during heating.
Our key findings are as follows:

{\it i)} The population of 6-membered rings reflects the relative stability of the glasses;
{\it ii)} Melting (transition from glassy to liquid state) and devitrification (transition from glassy to crystalline state) are observed at low and high heating rates, respectively, with the latter mediated by a transient supercooled liquid phase;
{\it iii)} Under instantaneous heating (effectively infinite rate), the melting kinetics of the QG follows the Avrami equation, while the DG exhibits a higher Avrami exponent, indicating enhanced kinetic stability;
{\it iv)} Melting in QG initiates in the bulk, whereas in DG it preferentially starts near the free surface, particularly under faster heating conditions;
{\it v)} Across all samples and heating rates, liquid clusters grow via sub-three-dimensional scaling, following the relation $n\sim R_g^{2.2}$, where $n$ is the cluster size and $R_g$ its radius of gyration;
{\it vi)} DG samples exhibit an excess of 5-membered rings near the free surface, consistent with their tendency to nucleate the crystal phase in this region—highlighting a distinct pathway compared to QG.

Overall, our results underscore the critical role of glass preparation protocol and interfacial effects in determining the thermal response and nucleation behavior of glassy water. The insights gained here may be relevant for understanding ultrastable glasses more broadly and for the design of materials with tailored thermal stability.

\section*{Data availability}

The data that support the findings of this study are available from the corresponding author upon reasonable request.


\begin{acknowledgements}
We are happy to join the celebrations for the 60th birthday of Carlos Vega.
We thank Andreas Neophytou for helpful discussions.
F.L. and J.R. acknowledge support by ICSC – Centro Nazionale di Ricerca in High Performance Computing, Big Data and Quantum Computing, funded by European Union – NextGenerationEU, and the CINECA award under the ISCRA initiative, for the availability of high-performance computing resources and support.
\end{acknowledgements}

\bibliography{biblio}

\end{document}